\documentclass[final,english]{bullsrsl}[2022/06/15]

\def\aap{A\&A}    
\def\apj{ApJ}     
\def\aarv{A\&ARv} 
\def\mnras{MNRAS} 
\def\apss{Ap\&SS} 
\def\LRSP{LRSP}   
\def\PEPS{PEPS}   

\usepackage[latin1]{inputenc}
\usepackage[T1]{fontenc}

\usepackage{natbib} 
\usepackage{graphicx}

\begin{document}
\title{Evolution of Solar-Type Activity: \\ An Observational and Theoretical Perspective}

\author[affil={}]{Manfred}{Cuntz}
\affiliation[]{Department of Physics, University of Texas at Arlington, Arlington, TX 76019, USA}
\correspondance{cuntz@uta.edu}
\date{13th October 2020}
\maketitle

\begin{abstract}
When stars depart from the main-sequence, various changes occur including the loss of angular momentum owing
to changes in the stellar interior and the impact of stellar winds.  These processes affect the amount of
outer atmospheric heating and emission as revealed by observations in the UV and X-ray spectral regimes.
From a theoretical perspective, both magnetic and acoustic energy generation are affected as indicated
by detailed theoretical simulations. Here, I will summarize selected observational and theoretical results,
including recent work for $\beta$~Hydri (G2~IV), a star constituting a prime example and proxy for the future Sun.
\end{abstract}

\keywords{stars: activity,
                stars: individual: $\beta$ Hydri,
                stars: magnetic fields,
                stars: chromospheres,
                stars: coronae,
                Sun: evolution}

\section{Introduction}

The study of the evolution of stellar activity is a central theme of astronomy and
astrophysics.  It is also of particular relevance to the Sun.  Important contributions
have been made about the relationships between the stellar age, mass loss, magnetic
field coverage, change in angular momentum, and outer atmospheric heating and emission
regarding the various spectral regimes.  Selected results will be highlighted in this paper
(see Section 2 for examples).

Aspects about solar-type evolution (sometimes also referred to as ``Sun in Time")
have been described in the living review by G\"udel (1997) and related studies.
The findings include that
(1) the Sun's magnetic activity has steadily declined during its main-sequence life,
associated with an increase in luminosity, 
(2) the young Sun exhibited significantly faster rotation generating enhanced magnetic activity,
(3) associated magnetic heating processes occurred in the chromosphere, the transition region,
and the corona; induced ultraviolet, extreme-ultraviolet, and X-ray emission has been about
10, 100, and 1000 times, respectively, higher compared to the present-day levels,
(4) drastic consequences happened regarding the Solar System (especially for close-in
planets and moons) due to high-energy particles and significant solar flares, and
(5) important notable processes occurred such as photoionization, photochemical evolution,
and possible erosion of the planetary atmospheres; the latter were highly relevant for
the origin or non-origin of life.

Some of these topics can be explored through the examination of solar analogs.
In the following, we will give special consideration to $\beta$~Hydri ($\beta$~Hyi;
HD 2151, HR 98, HIP 2021), a well-studied star in the Southern Hemisphere; see, e.g.,
Fawzy \& Cuntz (2023) for compiled information and references.  Previous work by
Brand{\~a}o et al. (2011) confirmed $\beta$~Hyi's spectral type as G2~IV, among
other properties.  The stellar parameters of $\beta$~Hyi read as follows:
$T_{\rm eff} = 5872$~K,
$M_\star=1.08~M_\odot$, $R_\star=1.81~R_\odot$, $L_\star=3.49~L_\odot$, and
${\log}~g_\star=4.02$~(cgs); see Da Silva et al. (2006), North et al. (2007),
Bruntt et al. (2010), and Brand{\~a}o et al. (2011); with all symbols having
their usual meaning.  The stellar age is given as $6.40 \pm 0.56$~Gyr; hence,
$\beta$~Hyi represents the medium-distant future of the Sun.

A highly comprehensive approach has been applied by Bruntt et al. (2010), who
combined results from interferometry, asteroseismology, and spectroscopy to derive
fundamental parameters of many stars, including $\beta$~Hyi.  Thus, the stellar
mass and effective temperature could be determined within an uncertainty of $\pm$0.05
and $\pm$59, respectively.  The method of Brand{\~a}o et al. (2011) relied on
asteroseismic modelling, including the consideration of seismic and non-seismic data
in conjunction with detailed evolutionary models.

Consequently, $\beta$~Hyi is of extraordinary interest to both
theoretical and observational studies, including the propagation and dissipation
of acoustic and magnetic energy in its outer atmosphere, as well as associated
chromospheric and coronal emission.  Previous observational studies allowing
to verify the status of $\beta$~Hyi employ the analysis of solar-like oscillations
(Carrier et al. 2001, Bedding et al. 2007).  Hence, $\beta$~Hyi is a valuable example
for solar studies in a broader context.  This paper is organized as
follows:  In Section 2, we present selected previous results on the
evolution of solar-type activity.  In Section 3, we discuss aspects of
$\beta$~Hyi.  Concluding remarks are given in Section 4.

\section{Selected Previous Results}

In the following, we present tidbits about the evolution of solar-type
activity pertaining to both observational and theoretical aspects.

\begin{itemize}

\item
A seminal paper was given by Noyes et al. (1984).  They
examined the relation between rotation, convection, and activity in low-mass
main-sequence stars.  Noyes et al. (1984) found that the mean levels of Ca~II H and K
chromospheric emission is well-correlated with the stellar rotational period,
in alignment with predictions from dynamo theory.  The authors also noted
a dependency of Ca~II H and K on the stellar spectral type that inspired many
future theoretical and observational studies.

\item
Schrijver (1987) and Rutten et al. (1991) studied sets of main-sequence stars
of different activity levels as well as giant stars while focusing on 
Ca~II H+K and Mg~II {\it h}+{\it k} chromospheric emission.  The authors
argue that the results indicate the presence of two-component chromospheric
emission, consisting of basal-flux emission (presumably of nonmagnetic, i.e.,
acoustic origin) and another type of emission that distinctly depends on
the stellar activity level.  The latter component is expected to be
magnetic in nature.  On the other hand, Judge \& Carpenter (1998) argued
in a subsequent study that the basal component is at least in part of
magnetic origin as well.

\item
Another important contribution was made by Baliunas et al. (1995), who
studied the Ca~II H+K emission cores in 111 stars of spectral type F2-M2
on or near the main-sequence.  The results indicate a pattern of change in
rotation and chromospheric activity on an evolutionary timescale, in which
(1) young stars exhibit high average levels of activity and rapid rotation rates;
(2) stars of intermediate age have moderate levels of activity and rotation rates,
and (3) stars as old as the Sun and older have slower rotation rates and lower
activity levels; the authors also commented on possible stellar Maunder minima.
More recent work by Wright et al. (2011) derived the stellar activity --- rotation
relationship for a sample of 824 solar and late-type stars, thus offering information
on the evolution of stellar dynamos.

\item
Charbonneau et al. (1997) studied various processes closely related
to the evolution of angular momentum based on detailed stellar structure
models including the change in inertia and the loss of angular momentum
through the stellar winds; moreover, the change of chromospheric
emission on stellar evolutionary time scales has been discussed as well.
Follow-up work has also been pursued, including studies by Keppens
et al. (1995), who explored the evolution of the rotational velocity
distribution for solar-type stars.  Furthermore, Saar \& Osten (1997)
studied the rotation, turbulence, and surface magnetic fields in
Southern dwarfs.

\item
Another relevant contribution was provided by P{\'e}rez Mart{\'{\i}}nez
et al. (2011).  They examined the Mg~II {\it h}+{\it k} line emission
associated with the extended chromospheres of a set of 177 cool G, K,
and M giants and supergiants.  These fluxes represent the chromospheric
radiative energy losses presumably related to basal heating by the
dissipation of acoustic waves, plus a highly variable contribution
due to magnetic activity.  The statistical analysis of the data
provide evidence for a well-defined Mg~II basal flux limit.

\item
Previous  examples for two-component (acoustic and magnetic) for
stellar chromosphere models allowing a detailed comparison with the
work by Schrijver (1987) and Rutten et al. (1991) have been given by
Cuntz et al. (1999).  They consider K2V stars of different
magnetic activity.  The magnetic filling factor (MFF) is set
via an observational relationship between the measured magnetic
area coverage and the stellar rotation period.  It is found that
the heating and chromospheric emission is significantly increased
in the magnetic component and is strongest for flux tubes that
spread the least with height, expected to occur on rapidly rotating
stars with high magnetic filling factors. For stars with very
slow rotation, the authors were able to reproduce the basal flux
limit previously identified with nonmagnetic regions.  The empirical
relationship between the Ca~II H+K emission and the stellar rotation
rate could be reproduced as well.  Additional results for stars
of other spectral types were given by Fawzy et al. (2002a, 2002b).
Previously, Buchholz et al. (1998) reproduced the basal flux limit
for both main-sequence stars and giants based on pure acoustic hearing.

\item
More recently, additional papers have been published for solar-type stars
pertaining to the relationships between stellar evolution, loss of
angular momentum, atmospheric heating and spectral emission,
including associated effects on exoplanets.  Examples include work
by Johnstone et al. (2015), Tu et al. (2015), and Linsky et al. (2020).
Johnstone et al. (2015) and Tu et al. (2015) focused on the evolution
of solar-type winds as well as the UV and X-ray behaviors of solar-type
stars on evolutionary time scales.
Linsky et al. (2020) pointed out, in alignment with previous studies,
that EUV and X-ray emission from stellar coronae drives mass loss from exoplanet
atmospheres, whereas emission from stellar chromospheres drives photochemistry
in exoplanet atmospheres.  This is a pivotal inspiration for further work,
including the study of the spectral energy distributions of host stars
--- noting that the latter is essential for understanding the evolution and
habitability of terrestrial exoplanets.  The authors concluded that as stars
age on the main-sequence, the emissions from their chromospheres and coronae
follow a pattern in response to the amount of magnetic heating in these
atmospheric layers.

\end{itemize}

\section{Case Study: Beta Hydri}

A key example for assessing the evolution of solar-type activity is $\beta$~Hyi.  It has
a surface temperature, metallicity, and mass akin to the Sun; however, it is $\sim$2 Gyr
older than the Sun (Dravins et al. 1998; Brand{\~a}o et al. 2011).  Hence, $\beta$~Hyi
constitutes a striking example of future solar activity and properties.  Previous
results have been given by, e.g., Dravins et al. (1993a,b,c).
These studies indicate that $\beta$~Hyi is at an early subgiant stage, when lithium
that once diffused to beneath the main-sequence convection zone is dredged up to the
surface as the convection zone deepens.  Other findings include that (1) according to
3-D photospheric hydrodynamic simulations, typical granules are about a factor of 5
larger than solar ones, (2) there is reduced photospheric pressure initiating
higher granular velocities, (3) high-resolution Ca II H and K profiles show the
emission to be about half that for the Sun, but with the same sense of violet-red
asymmetry, (4) weak UV and X-ray emission if compared to the Sun, and (5) stellar wind
conditions shaped by post--main-sequence thermodynamics.

However, most of these results have not yet been compared to detailed ab-initio models;
nonetheless, numerous findings about $\beta$~Hyi's outer atmospheric heating and
emission have become available based on theoretical simulations, including recent
studies by Cuntz \& Fawzy (2022) and Fawzy \& Cuntz (2023).  These models employ
a combination of acoustic and magnetic energy generation and propagation
(two-component models), including cases of different magnetic filling factors.
Previous theoretical work regarding magnetic and acoustic energy generation with a
focus on longitudinal flux-tube waves has been given by Musielak et al. (1994, 1995)
and Ulmschneider et al. (2001).  Specifics regarding two-component models has been
discussed by Cuntz et al. (1999) and others.

Results obtained by Cuntz and Fawzy include that for acoustic waves the amount of
generated energy is notably higher (about a factor of 2.3) in $\beta$~Hyi compared
to the Sun.  This behavior is due to the greater photospheric granular velocity of
$\beta$~Hyi owing to its lower surface gravity; see Dravins et al. (1993a).  For
longitudinal flux-tube waves, the amount of generated energy is reduced (about a
factor of 1.1 to 1.3) relative to the solar case, a behavior due to $\beta$~Hyi's
thermodynamic and magnetic field conditions.  Figure~1 conveys information about
the adopted flux tube models, whereas Figure~2 gives examples about the emergent
Ca~II core fluxes, both regarding $\beta$~Hyi and the Sun (note the impact of
${\log}~g_\star$!)  Smaller magnetic filling factors correspond to flux tubes
of more extended spreading accompanied with reduced heating and chromospheric
emission as a function of height.  The magnetic filling factor in $\beta$~Hyi
is expected to vary across the surface (just like in the solar case); therefore,
in the absence of detailed surface information, a set of models with different MFFs
should be taken into account.

The height-dependent behavior of the mechanical energy flux for different MFFs
regarding $\beta$~Hyi, as well as different types of models, is presented in
Figure~3; see Cuntz \& Fawzy (2022) for background information.  This work
considers both spectral and monochromatic waves.  It is found that
a higher MFF entails a somewhat smaller decrease of the magnetic wave
energy flux as a function of height, especially in the middle chromosphere,
mostly associated with the difference in tube spreading.  However, at
large heights other effects are also relevant, especially in narrow tubes,
including effects associated with strong shocks, which initiate both
quasi-adiabatic cooling.  The various kinds of models are a stark motivation
for more extended studies, including detailed comparisons between theoretical
and observational works.

\begin{figure}
\centering
\includegraphics{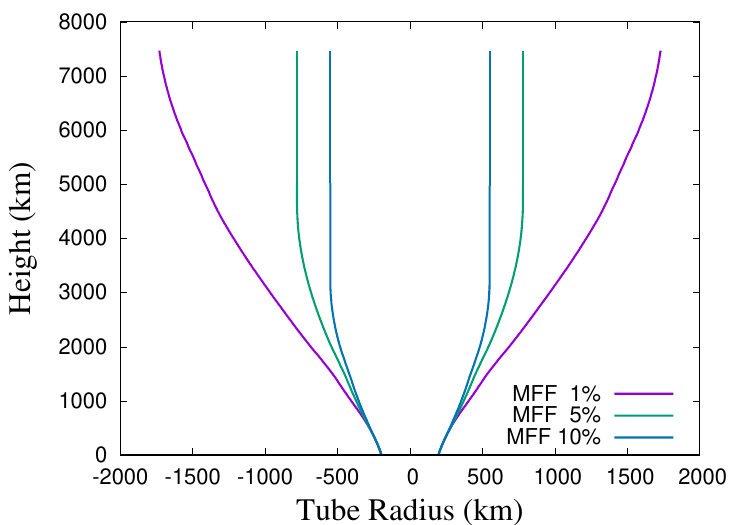}
\bigskip
\begin{minipage}{12cm}
\caption{
Flux tube models for $\beta$~Hyi based on different magnetic filling factors;
see Fawzy \& Cuntz (2023) for details.
}
\end{minipage}
\end{figure}

\begin{figure}
\centering
\includegraphics{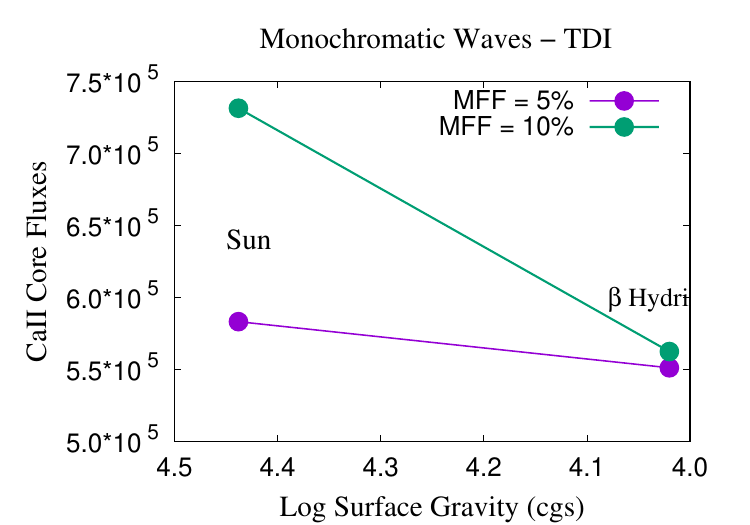}
\bigskip
\begin{minipage}{12cm}
\caption{
Emergent Ca~II core fluxes for two-component chromosphere models 
(acoustic and magnetic) pertaining to the Sun and $\beta$~Hyi
with photospheric magnetic filling factors given as 5\% and 10\%;
adopted from Cuntz \& Fawzy (2022).
}
\end{minipage}
\end{figure}

\begin{figure}
\centering
\includegraphics{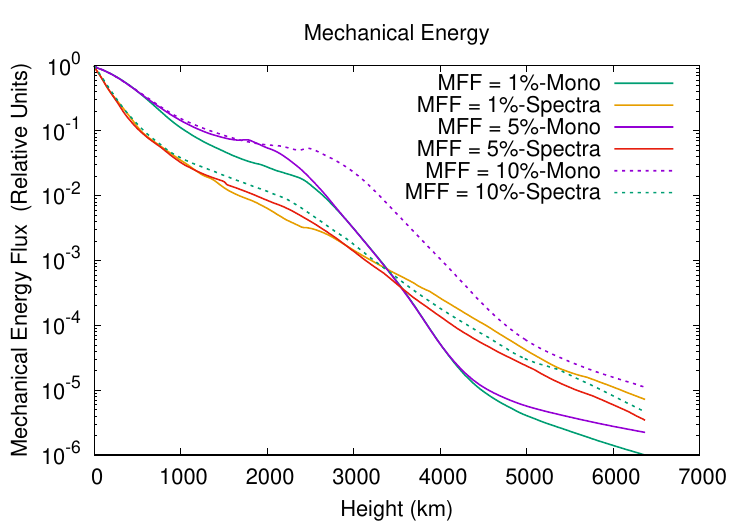}
\bigskip
\begin{minipage}{12cm}
\caption{
Height-dependent behavior of the mechanical energy flux in magnetic wave models
(relative units) for $\beta$~Hyi.  Both monochromatic waves and spectral waves
have been considered based on magnetic filling factors of 1\%, 5\%, and 10\%,
respectively; see Fawzy \& Cuntz (2023) for background information and details.
}
\end{minipage}
\end{figure}

\section{Concluding Remarks}

Future work is expected to provide more detailed comparisons between the outcomes
of theoretical models and observations in order to assess the modes of
outer atmospheric heating as well as the photospheric / chromospheric
magnetic filling factors.  The latter are a distinct testimony of stellar evolution,
which is closely connected to the loss of angular momentum resulting in a reduction in
magnetic activity.  In case of the Sun, its activity varies on different timescales,
including very short and super-long timescales, with the latter being associated with
stellar and planetary evolution (e.g., Nandy et al. 2021, and references therein).
Clearly, those latter changes are closely related to the Sun's nuclear evolution,
thus affecting the solar interior structure, which continues to result
in the loss of angular momentum as well as changes in the solar rotation rate,
magnetic surface structure, and the magnitude of magnetic and nonmagnetic surface
heating and outer atmospheric emission.

For example, the environment of the early, young Sun was relatively extreme, with
more energetic storms, winds, and radiative output creating harsh conditions for
(especially) the inner Solar System planets, including Earth.  These conditions will
continue to abate, which will favorably affect the habitability prospects of other
Solar System objects, including the terrestrial satellites of Jupiter and Saturn.
Reviews on stellar activity and its role(s) on space weather have been given by
Lammer et al. (2009) and others; see also Whitepaper by
Airapetian et al. (2018) and subsequent contributions.
This type of work is highly relevant to a large array of topics including
(but not limited to) the future of life on Earth, general prospects of planetary
habitability concerning different types of stars, and the search for life in the Universe.

\begin{acknowledgments}
The author is grateful to the scientific and local organizing committees
of BINA-3 for the invitation to this workshop.  In addition, he appreciates
travel support from The University of Texas at Arlington.
\end{acknowledgments}

\begin{furtherinformation}

\begin{orcids}
\orcid{0000-0002-8883-2930}{Manfred}{Cuntz}
\end{orcids}

\begin{conflictsofinterest}
The author declares no conflict of interest.
\end{conflictsofinterest}

\end{furtherinformation}



\newpage

\noindent
{\bf References}

\medskip

\noindent
Airapetian, V.S., Adibekyan, V., Ansdell, M., et al.: arXiv \ 1803.03751 (2018)

\noindent
Baliunas, S.L., Donahue, R.A., Soon, W.H., et al.:
\apj \ {\bf 438,} 269 (1995)

\noindent
Bedding, T.R., Kjeldsen, H., Arentoft, T., et al.:
\apj \ {\bf 663,} 1315 (2007)

\noindent
Brand{\~a}o, I.M., Do{\u{g}}an, G., Christensen-Dalsgaard, J., et al.:
\aap \ {\bf 527,} A37 (2011)

\noindent
Bruntt, H., Bedding, T.R., Quirion, P.-O., et al.: \mnras \ {\bf 405,} 1907 (2010)

\noindent
Buchholz, B., Ulmschneider, P., Cuntz, M.: \apj \ {\bf 494,} 700 (1998)

\noindent
Carrier, F., Bouchy, F., Kienzle, F., et al.: \aap \ {\bf 378,} 142 (2001)

\noindent
Charbonneau, P., Schrijver, C.J., MacGregor, K.B.: In:
Jokipii, J.R., Sonett, C.P., Giampapa, M.S. (eds.),
Cosmic Winds and the Heliosphere. Space Science Series,
Univ. of Arizona Press, Tucson, 677 (1997)

\noindent
Cuntz, M., Fawzy, D.E.: RNAAS \ {\bf 6,} 147 (2022)

\noindent
Cuntz, M., Rammacher, W., Ulmschneider, P., Musielak, Z.E., Saar, S.H.: \apj \ {\bf 522,} 1053 (1999)

\noindent
Da Silva, L., Girardi, L., Pasquini, L., et al.: \aap \ {\bf 458,} 609 (2006)

\noindent
Dravins, D., Lindegren, L., Nordlund, {\AA}, VandenBerg, D.A.: \apj \ {\bf 403,} 385 (1993a)

\noindent
Dravins, D., Linde, P., Fredga, K., Gahm, G.F.: \apj \ {\bf 403,} 396 (1993b)

\noindent
Dravins, D., Linde, P., Ayres, T.R., et al.: \apj \ {\bf 403,} 412 (1993c)

\noindent
Dravins, D., Lindegren, L., VandenBerg, D.A.: \aap \ {\bf 330,} 1077 (1998)

\noindent
Fawzy, D.E., Cuntz, M.: \apss \ {\bf 368,} 1 (2023)

\noindent
Fawzy, D., Rammacher, W., Ulmschneider, P., Musielak, Z.E., St{\c e}pie{\'n}, K.: \aap \ {\bf 386,} 971 (2002a)

\noindent
Fawzy, D., St{\c e}pie{\'n}, K., Ulmschneider, P., Rammacher, W., Musielak, Z.E.: \aap \ {\bf 386,} 994 (2002b)

\noindent
G\"udel, M.: \LRSP \ {\bf 4,} 3 (1997)

\noindent
Johnstone, C.P., G\"udel, M., Brott, I., L\"uftinger, T.: \aap \ {\bf 577,} A28 (2015)

\noindent
Judge, P.G., Carpenter, K.G.: \apj \ {\bf 494,} 828 (1998)

\noindent
Keppens, R., MacGregor, K.B., Charbonneau, P.: \aap \ {\bf 294,} 469 (1995)

\noindent
Lammer, H., Bredeh\"oft, J.H., Coustenis, A., et al.: \aarv \ {\bf 17,} 181 (2009)

\noindent
Linsky, J.L., Wood, B.E., Youngblood, A., et al.: \apj \ {\bf 902,} 3 (2020)

\noindent
Musielak, Z.E., Rosner, R., Stein, R.F., Ulmschneider, P.: \apj \ {\bf 423,} 474 (1994)

\noindent
Musielak, Z.E., Rosner, R., Gail, H.P., Ulmschneider, P.: \apj \ {\bf 448,} 865 (1995)

\noindent
Nandy, D., Martens, P.C.H., Obridko, V., Dash, S., Georgieva, K.:
\PEPS \ {\bf 8,} 40 (2021)

\noindent
North, J.R., Davis, J., Bedding, T.R., et al.: \mnras \ {\bf 380,} L80 (2007)

\noindent
Noyes, R.W., Hartmann, L.W., Baliunas, S.L., Duncan, D.K., Vaughan, A.H.: \apj \ {\bf 279,} 763 (1984)

\noindent
P{\'e}rez Mart{\'{\i}}nez, M.I., Schr\"oder, K.-P., Cuntz, M.: \mnras \ {\bf 414,} 418 (2011)

\noindent
Rutten, R.G.M., Schrijver, C.J., Lemmens, A.F.P., Zwaan, C.: \aap \ {\bf 252}, 203 (1991)

\noindent
Saar, S.H., Osten, R.A.: \mnras \ {\bf 284,} 803 (1997)

\noindent
Schrijver, C.J.: \aap \ {\bf 172}, 111 (1987)

\noindent
Tu, L., Johnstone, C.P., G\"udel, M., Lammer, H.: \aap \ {\bf 577}, L3 (2015)

\noindent
Ulmschneider, P., Musielak, Z.E., Fawzy, D.E.: \aap \ {\bf 374,} 662 (2001)

\noindent
Wright, N.J., Drake, J.J., Mamajek, E.E., Henry, G.W.:
\apj \ {\bf 743,} 48 (2011)


\end{document}